\documentclass[aps,prl,twocolumn,showpacs,superscriptaddress,groupedaddress]{revtex4}  
\usepackage{graphicx}  
\usepackage{dcolumn}   
\usepackage{bm}        
\usepackage{amssymb}   
\usepackage{amsmath}
\usepackage{cancel}
\usepackage{empheq}
\usepackage{graphicx,color}

\begin{document}


\title{Ring closure in actin polymers}
\author{Supurna Sinha $^{}$ }
\address{Raman Research Institute, Bangalore 560080, India.}
\author{Sebanti Chattopadhyay $^{}$}
\address{Doon University, Dehradun 248001, India.}
\date{\today}

\begin{abstract}

We present an 
analysis for the ring closure probability of semiflexible polymers within the pure bend Worm Like Chain (WLC) model. 
The ring closure probability 
predicted from our analysis can be tested 
against fluorescent actin cyclization experiments.{We also discuss the effect of ring closure on bend angle fluctuations in actin polymers.} 

\end{abstract}

\maketitle


\section{{\bf{I.}} Introduction}
In the past two decades, there has been much interest in the theoretical study of
semiflexible polymer elasticity. These studies are motivated by
micromanipulation experiments\cite{bust,sethnawang,sethnadaniels}
on biopolymers. {In particular, in recent years there have been 
experiments involving stretching DNA molecules\cite{bust} which give us information about the bend elastic properties of DNA. 
There have also been experiments on fluorescently tagged actin filaments\cite{lib} where they measure the bend persistence length 
of actin. More recently, there have been  
fluorescence experiments on 
cyclization of actin filaments\cite{actincirc}.  
In these papers they analyze the formation of rings in actin polymers and study the effect of ring closure on 
bend angle fluctuations in these polymeric rings.}
Our interest here is
limited to the process of cyclization itself and therefore in our analysis 
we restrict to polymers with only bend degrees of freedom and no twist degree of freedom. Actin cyclization is
of interest to biologists \cite{actnat} who do visualization studies of actin ring formation in the context of 
cell division.    


\section{{\bf{II.} }Ring Closure Probability Distribution}
Our starting point is the pure bend Worm Like 
Chain (WLC) model\cite{js}. 
In this model, the polymer configuration is viewed as a space curve
${\vec x(s)}$. There is a tangent vector associated with each point 
on the polymer of contour length $L$ and the energy of configuration is given by:
\begin{equation}
{\cal {E[C]}}=\frac{A}{2}\int_0^L{ds {\kappa}^2}
\label{energy}
\end{equation} 
where $\cal{C}$ stands for the polymer configuration. $A$ is the bending elastic constant and the curvature $\kappa = |\frac{d{\hat{t}}}{ds}|$.

One of the key quantities characterizing the elasticity of a biopolymer is 
{$\tilde{Q}(\vec r)$}, the probability distribution for the end to end distance vector $\vec r$ between the two ends of 
the polymer as it gets jiggled around by thermal fluctuations in a cellular environment \cite{js}. 
In \cite{js} we use a method for solving the wormlike chain model for semiflexible polymers to any desired
accuracy over
the entire range of polymer lengths to determine {$\tilde{Q}(\vec r)$}. The plots for {$\tilde{Q}(\vec r)$} for various $\beta=\frac{L}{L_P}$,
the ratio of the contour length $L$ to the persistence length $L_P$, reveal
the dependence of the end to end distance vector on the rigidity of the polymer (See Fig. [4] in \cite{js}). 

We outline the theoretical calculation of {$\tilde{Q}({\vec r})$} below. {(For a detailed exposition please see 
Appendix $A$).} Consider a situation where the initial and 
final tangent vectors (${\hat t}_A=\frac{d {\vec x}}{ds}|_{s=0}$ and ${\hat t}_B=\frac{d {\vec x}}{ds}|_{s=L}$)
are held fixed. Then {$\tilde{Q}({\vec r})$} has the following path integral representation:
\begin{eqnarray}\nonumber
{\tilde{Q}({\vec r})}&=& {\cal N}\int{{\cal D}[{\hat t(s)}] exp\{{-\frac{1}{k_BT}\big[\frac{A}{2}\int_0^L{(\frac{d {\hat t}}{ds})^2 ds}\big]\}}}\\
& &{\times \delta^{3}(\vec{r}-\int_0^L{{\hat t} ds})  } 
\label{qofr}
\end{eqnarray}
Here ${\cal N}$ is the normalization constant and $k_BT$, the thermal energy at temperature $T$. 
As mentioned in \cite{js}, we solve for {$\tilde{Q}(\vec r)$} by first considering a related end to end distance measure: 
{$$P(z)=\int{d{\vec r} {\tilde{Q}({\vec r})} \delta (r_3 -z)},$$} 
which is {$\tilde{Q}(\vec r)$} integrated over a plane of constant $z$. 
This in turn is related to ${\tilde P}(f)$, the Laplace transform
of $P(z)$ given by:
\begin{eqnarray}
{\tilde P}(f)=\int_{-L}^L{P(z) e^{\frac{fz}{L_P}}dz}
\label{laplace}
\end{eqnarray}
$f$, the variable conjugate to $z$ has the interpretation of a 
stretching force and thus ${\tilde P}(f)$, can be written as the ratio 
$Z(f)/Z(0)$ of the 
partition functions in the presence and absence of an external stretching force $f$. 
We do
an eigenspectrum analysis of 
${\tilde P}(f)$ and determine {$\tilde{Q}(\vec r)$} using tomographic transformations outlined in 
\cite{js}. 

Here we address a question which is of current interest to application of polymer physics to biology: cyclization of 
actin filaments\cite{actincirc}. 
Within the pure bend Worm Like Chain (WLC) Model we compute the ring closure probability (RCP) by 
considering {$\tilde{Q}(\vec r=\vec{0})$}.   

\section{{\bf{III.}} Method}
In Fig. $4$ of \cite{js} we display a family of curves of $Q(\rho)$ versus $\rho$, with $\rho=\frac{|{\vec r}|}{\beta}$
for various values of $\beta$. {$Q(\rho)$ is a theoretically convenient quantity expressed in terms of scaled units 
($\vec{\rho}=\frac{{\vec r}}{\beta}$).}
In order to compute the ring closure probability density {$\tilde{Q}(\vec{r}=\vec{0})$} 
we need to change variables from $\rho=\frac{|{\vec r}|}{\beta}$ 
to $|{\vec r}| = r$. Setting {$\tilde{Q}({\vec r})=Q_{\vec r}$}, we get: 

\begin{eqnarray}
\int{Q({\vec \rho})}d{\vec \rho}&=& \int{Q_{\vec r}}d{\vec r}   
\label{qofzero}
\end{eqnarray}

or

\begin{eqnarray}
\int{\frac{Q({\vec \rho})}{\beta^3}d{\vec r}}&=& \int{{Q_{\vec r}}d{\vec r}}   
\label{qofzeror}
\end{eqnarray}

which in turn implies

\begin{eqnarray}
{\frac{Q(0)}{\beta^3}}&=& {Q_0}   
\label{cov}
\end{eqnarray}
We compute $Q(0)$ for a range of values of $\beta$ using Mathematica.
{As we can see from the plot of the ring closure probability density $Q(0)$ versus $\beta$ (Fig. $1$),
that $Q(0)$ has a small value for short polymers which are hard
to bend and form rings and it has a large value for long polymers which are
easy to bend and thus the probability density of ring formation is high.
We then compute and} plot the ring closure probability density in physical space, $Q_0=\frac{Q(0)}{\beta^3}$ as a function of $\beta$ (Fig. {$2$}). 
The qualitative features of the plot shown in Fig. {$2$} are in agreement with our intuition. The ring closure probability density
$Q_0$ in physical space, which is an experimentally measurable quantity is small for very short and long strands of the polymer and peaks around intermediate contour lengths 
of $L \approx 3 L_P$ {(See Fig. 7-41 on page 438 of \cite{cell}).}

\begin{figure}
\includegraphics[scale=0.5]{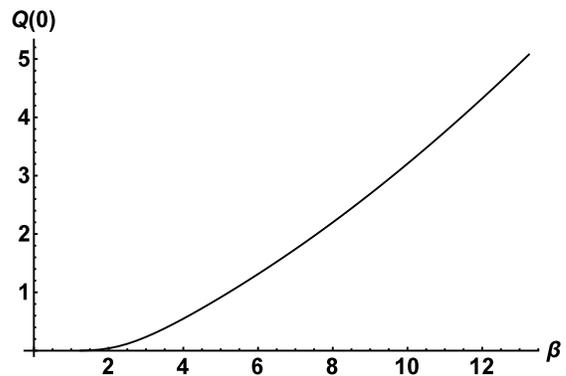}
\caption{\label{fig:epsart} 
{A plot of the ring closure probability density $Q(\vec{\rho}=\vec{0})=Q(0)$ versus $\beta$, setting $L_P=1$. 
It has a small value for short polymers which are hard
to bend and form rings and it has a large value for long polymers which are
easy to bend and thus the probability density of ring formation is high.}}
\end{figure}

\begin{figure}
\includegraphics[scale=0.5]{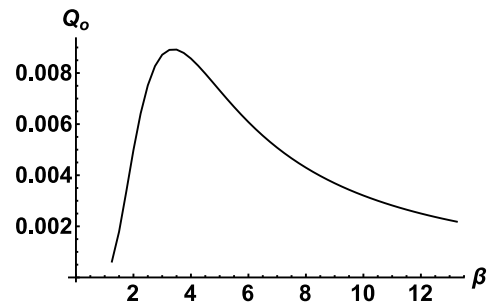}
\caption{\label{fig:epsart} 
{A plot of the ring closure probability density in physical space  
$Q_0=Q(0)/\beta^3$, versus $\beta$ {setting $L_P=1$}. Notice that this function is small for very small and large
$\beta$ and peaks around an intermediate value $\beta \approx 3$. }}
\end{figure}

\begin{figure}
\includegraphics[scale=0.5]{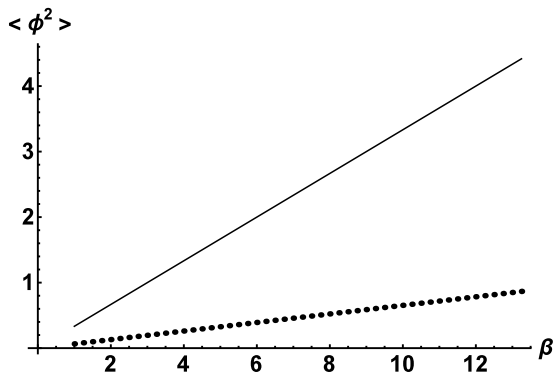}
\caption{\label{fig:epsart} 
{Plots of mean squared tangent angle fluctuation in three dimensions for a ring (dashed line) and a linear filament (solid line).
We have set $L_P=1$.
Notice the suppression of fluctuation for a ring filament compared to a linear one. }}
\end{figure}

{\section{\bf{IV}. Mean squared tangent angle fluctuation}
One of the experimentally relevant quantities of interest is the mean squared tangent angle fluctuation\cite{actincirc}.
In Ref. \cite{actincirc} the mean squared tangent angle fluctuation has been calculated for a ring and a linear filament in a two dimensional setup. They find good agreement with experimental measurements. 

Here we present a similar calculation in a three dimensional geometry. 
Consider a polymer configuration in a closed circular ring lying in the $x-y$ plane. 
Expanding the bend angle fluctuation $\phi(s)$ 
in a Fourier series and imposing the ring closure constraint and removing zero modes which do not contribute, we find that the contribution from the $x-y$ plane is given by
\begin{equation}
{<\phi^2>^{xy}_{ring}=\frac{1}{12}(1-\frac{6}{\pi^2})\frac{L}{L_P}}
\label{phi2}
\end{equation}
We need to add this contribution to the contribution coming from the $z$ direction where the ring closure condition is of the form
$$\int_0^L{\phi_z(s) ds}=0.$$

In this case the Fourier expansion for $\phi_z(s)$ can be expressed as $\phi_z(s)=\sum_{n=2}^{\infty}{\phi_n{e^\frac{2\pi i n s}{L}}}$
which finally gives us 
\begin{equation}
{<\phi^2>^{z}_{ring}=\frac{1}{12}(1-\frac{6}{\pi^2})\frac{L}{L_P}}
\label{phiz2}
\end{equation}

Thus combining Eqs. \ref{phi2} and \ref{phiz2}, the net mean squared tangent angle fluctuation for a three dimensional ring is given by
\begin{equation}
{<\phi^2>^{3d}_{ring}=\frac{1}{6}(1-\frac{6}{\pi^2})\frac{L}{L_P}}
\label{phi2ring}
\end{equation}
A similar calculation for a linear filament in three dimensions gives us
\begin{equation}
{<\phi^2>^{3d}_{lin}=\frac{1}{3}\frac{L}{L_P}}
\label{phi2line}
\end{equation}
We have plotted (\ref{phi2ring}) and (\ref{phi2line}) in Fig. $3$. These predictions can be tested against future experiments on fluorescently 
tagged actin filaments. Notice that, as in the two dimensional case\cite{actincirc}, we find that $<\phi^2>$ is suppressed for a ringlike structure compared
to a linear filament. 
This indicates that cyclization is entropically costly. 
Also, as expected, the fluctuations are smaller in the two dimensional geometry compared to the three dimensional 
geometry.}

\section{\bf{V}. Conclusion}
Our treatment is an 
analysis based on the pure bend Worm Like Chain Model. The absence of 
twist degree of freedom enables our predictions to be directly tested against actin cyclization
experiments where the two ends of the polymer come together without any relative twist 
between the two ends. This is to be contrasted with analysis of J factor of DNA with twist degree of freedom 
where the additional twist degree of freedom makes the analysis considerably more cumbersome\cite{sstwist,ajspakowitz,yama}. 
In \cite{yama} an interpolation formula is presented in the intermediate rigidity regime.
{They \cite{yama} also presented analytical expressions for the ring closure probability density $Q_0$ in the limit of $\beta<<1$ and $\beta>>1$.
However, they did not have an exact expression for the entire range of polymer lengths. 
In contrast, here we present a semianalytical study  which gives an essentially exact prediction for $Q_0$ 
for the {\it{entire}} range of rigidity (See Fig. $1$ and Fig. $2$).}   
It would be interesting to 
see how our predictions quantitatively compare with future cyclization probability data for actin filaments. 
{We also expect our predictions for the mean squared tangent angle fluctuation to be tested against future experiments on 
fluorescently tagged ring like and linear actin filaments in a three dimensional geometry.

}
\section{Acknowledgement}
One of us (SC) would like to thank RRI for providing hospitality during stay in 
Raman Research Institute as a Visiting Student.


\section{Appendix A: Computation of $Q_0$}

Our goal is to calculate $Q_0$, the ring closure probability. $Q_0$ is $\tilde{Q}(\vec{r})$ for $\vec{r}=\vec{0}$. 
$\tilde{Q}(\vec{r})$, the probability distribution for the end to end vector $\vec{r}$ for a semiflexible polymer has 
the following path integral representation:
\begin{eqnarray}\nonumber
\tilde{Q}({\vec r})&=& {\cal N}\int{{\cal D}[{\hat t(s)}] exp\{{-\frac{1}{k_BT}\big[\frac{A}{2}\int_0^L{(\frac{d {\hat t}}{ds})^2 ds}\big]\}}}\\
& &{\times \delta^{3}(\vec{r}-\int_0^L{{\hat t} ds})  }
\label{qofr}
\end{eqnarray}
Here ${\cal N}$ is the normalization constant and $k_BT$, the thermal energy at temperature $T$.

Instead of $\tilde{Q}(\vec{r})$ it turns out to be easier to first consider $P(z)$
$$P(z)=\int{d{\vec r} \tilde{Q}({\vec r}) \delta (r_3 -z)},$$
which is $\tilde{Q}({\vec r})$ integrated over a plane of constant $z$. 

$P(z)$ in turn is related to ${\tilde P}(f)$, the Laplace transform
of $P(z)$ given by:
\begin{eqnarray}
{\tilde P}(f)=\int_{-L}^L{P(z) e^{\frac{fz}{L_P}}dz}
\label{laplace}
\end{eqnarray}
$f$, the variable conjugate to $z$ has the interpretation of a
stretching force and thus ${\tilde P}(f)$, can be written as the ratio
$Z(f)/Z(0)$ of the
partition functions in the presence and absence of an external stretching force $f$.

$Z(f)$ is given by

\begin{eqnarray*}
Z(f) = {\cal N} \int {\cal D} \left[\hat{t}(s)\right] \exp \left\{ -\frac{L_{P}}{2} \left[ \int^{L}_{o} \left (\frac{d\hat{t}}{ds}\right)^{2} ds\right]\right\} \\
\times\exp \left[\frac{f}{L_{P}} \int^{L}_{0} \hat{t}_{2} ds\right] 
\end{eqnarray*}
which in turn can be expressed as 

\begin{eqnarray*}
Z(f) = {\cal N} \int {\cal D} \left [\hat{t}(\tau)\right] \exp \left\{ - \int^{\beta}_{o} d\tau \left[\frac{1}{2} \left(\frac{d\hat{t}}{d\tau}\right)^{2} - f\hat{t}_{z}\right]\right\} 
\end{eqnarray*}
which has the interpretation of the kernel of a quantum particle on the surface of a sphere at an inverse temperature $\beta$.
We now exploit the analogy between time imaginary quantum mechanics and classical statistical mechanics to re-express $Z(f)$ as 
follows:

\begin{eqnarray*}
Z(f) = \sum_{n} e^{-[\beta E_{n}]} \psi_{n} \left(\hat{t}_{A}\right) \psi_{n} \left(\hat{t}_{\beta}\right)  
\end{eqnarray*}

where $\{\psi_{n} \left(\hat{t}_{A}\right)\}$ is a complete set of normalized eigenstates of the Hamiltonian
$H_{f} = - \frac{\nabla^{2}}{2} - f \cos \theta$ and $E_n$ are the corresponding eigenstates.  
For free boundary conditions for the end tangent vectors we can express $Z(f)$ as 
\begin{eqnarray*}
Z(f) = \left< o \left| \exp - \beta H_{f} \right| o\right>   
\end{eqnarray*}
The Hamiltonian $H_{f} = - \frac{\nabla^{2}}{2} - f \cos \theta$   
is the Hamiltonian of a rigid rotor in a potential and $|0>$ is the ground state of the free Hamiltonian 
$H_0 = - \frac{\nabla^{2}}{2} $. We numerically evaluate $Z(f)$ by choosing a suitable basis in which $H_f$ has a tridiagonal symmetric 
matrix structure with 

$$H_{ll} = \frac{l(l+1)}{2}$$   

$$H_{ll+1} = f (l+1) \sqrt{1/[(2l + 1) (2l +3)}]$$   

To summarize, after casting the problem analytically we use Mathematica programs to sequentially compute 
$Z(f)$ and ${\tilde P}(f)$, then $P(z)$, then $S(r) = -2 r dP(r)/dr =4\pi r^{2} \tilde{Q} (r)$ and finally $\tilde{Q}(\vec{r})$. 
We then consider the scaled variable $\vec{\rho}=\frac{\vec{r}}{\beta}$. $Q(0)$ is then computed by considering
$Q(\vec{\rho})$ at $\vec{\rho}=\vec{0}$ and plotting it as a function of $\beta$. 
$Q_0=\frac{Q(0)}{\beta^3}$. Below we display the programs for computing $Q(\vec{\rho})$ and $Q_0$.  
We have inserted some comments as part of the programs for clarity. 

\section*{Program for computing $Q(\vec{\rho})$}

ClearAll[h,f,Z,lmax,H,beta,L1,L2,LPR,PR]\\
lmax=10;\\
Nmax=3000;\\
beta=3;\\
h=.005;\\
L=\{\};\\
For[$n=0$,$n<Nmax+1$,n++,\\
f=h*n*I;\\
$H=Table[Switch[i-j,-1,f*(i+1)/Sqrt[\\
(2i+1)(2i+3)],0,i (i+1)/2,1,f*(i)/Sqrt[(2i-1)(2i+1)],_,0],{i,0,lmax},{j,0,lmax}]$;\\
M=MatrixExp[-beta*H];\\
(*Computation of Z(f)*)\\
Z=M[[1,1]];\\
L=Append[L,Z]]\\
L=Re[L];\\
Pz=\{\};\\
P1z=\{\}\\
For[l=-2,l<1200,l++,\\
xi=.001*l;\\
P=(h*beta/Pi)*Sum[L[[n]]*Cos[(n-1)*h*xi*beta],{n,1,Nmax}];\\
Pz=Append[Pz,{xi,P}];\\
P1z=Append[P1z,P]];\\
V=P1z;\\
QR1={};\\
L1=Drop[V,2]\\
L2=Drop[V,-2]\\
LPR=(L1-L2)/(.001*2);\\
LPR=Drop[LPR,1];\\
(*Computation of S(r)*)\\
$QR=Table[LPR[[i]]*1/((i-1)*.001)*[-1/(2*Pi)],{i,2,1199}]$;\\
(*Computation of Q(r)*)\\
$QR1=Table[{(i-1)*.001,LPR[[i]]*(1/(i-1))}*\\
{1/((i-1)*.001)*[-1/(2*Pi)]},{i,2,2}]$;\\
ListPlot[QR1]\\

\section*{Program for computing $Q_0$}
ClearAll[h, f, Z, lmax, H, beta, Nmax, L1, L2, LPR, PR]\\
lmax = 10;\\
h = .005;\\
final = \{\};\\
For[k = 0, $k < 50$, k++,\\
beta = .25*k + 1;\\
Nmax = Piecewise[{{90000, beta $<= 3$}, {9000, beta $> 3$}}];\\
L = \{\};\\
For[n = 0, $n < Nmax + 1$, n++, f = h*n*I;\\
$H=Table[Switch[i-j,-1,f*(i+1)/Sqrt[\\
(2i+1)(2i+3)],0,i (i+1)/2,1,f*(i)/Sqrt[(2i-1)(2i+1)],_,0],{i,0,lmax},{j,0,lmax}]$;\\
M = MatrixExp[-beta*H];\\
Z = M[[1, 1]];\\
L = Append[L, Z]]\\ 
L = Re[L];\\
Pz = \{\};\\
P1z = \{\};\\
For[$l = -2, l < 1200, l++, xi = .001*l$;\\
P = (h*beta/Pi)*Sum[L[[n]]*Cos[(n - 1)*h*xi*beta], {n, 1, Nmax}];\\
Pz = Append[Pz, {xi, P}];\\
P1z = Append[P1z, P]];\\
V = P1z;\\ 
QR1 = \{\};\\
L1 = Drop[V, 2];\\
L2 = Drop[V, -2];\\
LPR = (L1 - L2)/(.001*2);\\
LPR = Drop[LPR, 1];\\
$QR=Table[LPR[[i]]*1/((i-1)*.001)*[-1/(2*Pi)],{i,2,1199}]$;\\
$QR1=Table[{(i-1)*.001,LPR[[i]]}*\\
{(1/(i-1))*1/((i-1)*.001)*[-1/(2*Pi)]},{i,2,2}]$;\\
$final = Append[final, {beta, (1/beta^3)*QR[[1]]}]$]\\
ListPlot[final]\\


\end{document}